\documentclass[twocolumn,showpacs,prl]{revtex4}

\usepackage{amsmath}

\usepackage{amssymb}
\begin{document}


\title{Entanglement of formation
for an arbitrary two-mode Gaussian state}
\author{Paulina Marian}
\author{ Tudor A. Marian}
\affiliation{ Centre for Advanced  Quantum Physics,
University of Bucharest, P.O.Box MG-11,
R-077125 Bucharest-M\u{a}gurele, Romania}
\date{\today}

\begin{abstract}
We write the optimal pure-state decomposition of any two-mode
Gaussian state and show that its entanglement of formation
coincides with the Gaussian one. This enables us to develop
an insightful approach of evaluating the exact entanglement
of formation. Its additivity is finally proven.

\end{abstract}


\maketitle
In recent years impressive efforts have been made
to quantify the entanglement of two-party states
of quantum systems. This trend was highly stimulated
by the interest in exploiting entanglement as an efficient
resource in quantum information processing. For any pure
bipartite state a convenient measure of entanglement
is now unanimously admitted, namely, the von Neumann entropy
of its reduced states \cite{BBPS,Ved}. Unlike the pure-state
case, several measures of entanglement have been
considered for mixed bipartite states on both finite- and
infinite-dimensional Hilbert spaces \cite{Plenio}.
Because of its operational meaning, the {\em entanglement
of formation} (EF) of a mixed bipartite state,
introduced by Bennett {\em et al.} \cite{Ben}, plays
a significant role: it is the minimal amount of entanglement
of any ensemble of pure bipartite states realizing the given
state. To be explicit, the EF of a mixed bipartite state $\rho$
is defined as an infimum  taken over all its pure-state convex
decompositions \cite{Ben}:
\begin{eqnarray}
E_F(\rho):=\inf \{\sum_{k}p_k
E(|\Psi_k\rangle\langle\Psi_k|)\mid
\rho=\sum_{k}p_k|\Psi_k\rangle\langle\Psi_k| \}.  \label{EF}
\end{eqnarray}
Here
$E(|\Psi_k\rangle\langle\Psi_k|)$ is the amount of entanglement
of the pure bipartite state $|\Psi_k \rangle.$ According to
definition\ (\ref{EF}), evaluating the EF is a hard task,
even for special quantum states. However, analytic evaluations
of the EF have been carried out in a few finite-dimensional cases:
general two-qubit states \cite{Wootters}, isotropic states
\cite{Terhal}, and Werner states \cite{Vollbrecht}.

In quantum information with continuous variables, two-mode
Gaussian states (TMGSs) of the quantum radiation field
are especially accessible from both theoretical and experimental
 standpoints.
Their usefulness  was recently reviewed
in Refs. \cite{EPBL,FAI}. So far, the only evaluation of the
exact EF in an infinite-dimensional Hilbert space has been
performed for symmetric TMGSs \cite{Giedke}. Moreover,
the additivity of the EF has been proven in this case
\cite{Wolf}. The Peres-Simon separability theorem \cite{Simon}
made it possible to use Gaussian measures of entanglement.
Within a Gaussian approach, the reference set of states involved
in the definition of any accepted entanglement measure
is restricted to the subset of the Gaussian ones. Thus, following
an earlier distance-type proposal for quantifying entanglement
due to Vedral and co-workers \cite{Ved}, several Gaussian
evaluations employing the relative entropy \cite{Scheel}
or the Bures metric \cite{PTH1,PT2,PT3} have been performed.
In Ref.\cite{Wolf}, a {\em  Gaussian entanglement
of formation} (GEF) has been introduced for any inseparable
TMGS by analyzing its optimal decomposition into pure TMGSs.

The aim of the present work is threefold. First, we build the appropriate decomposition, Eq.\ (\ref{EF}), of a TMGS $\rho_G$
that allows us to show that its EF and  GEF coincide.
We thus answer an open problem in continuous-variable quantum information \cite{Giedke,Wolf}.
Second, we give a more comprehensible approach to the problem
of evaluating the GEF by use of covariance matrices (CMs).
This enables us to write equations that yield, via the resulting optimal decomposition, an analytic solution for the EF
in the general case. We also get explicit results in the most interesting special cases.
Third, based on this approach, we prove the additivity of the EF
for two-mode Gaussian states.

Before proceeding we recall several useful properties of TMGSs.
For later convenience, we choose to describe any TMGS  $\rho_G$ by
its characteristic function (CF),
\begin{equation}
\chi_G(\lambda_1,\lambda_2):= {\rm Tr} [\rho_G D_1(\lambda_1)D_2(\lambda_2)],\end{equation}
where  $D(\alpha):=\exp{(\alpha a^{\dag}-\alpha^* a)}$ is a Weyl
displacement operator. The CF of an undisplaced TMGS is
$\chi_G(x)=\exp{\left(-\frac{1}{2}x^T {\cal V} x \right)}$. Here
$x \in \mathbb{R}^4$  and
${\cal V}$ is the real, symmetric, and positive $4\times 4$
CM that completely describes the state. Its entries are
the second-order moments of the canonical operators
$q_j=(a_j+a_j^{\dag})/{\sqrt{2}},\; p_j=(a_j-a_j^{\dag})/(\sqrt{2}i)$,
where $a_j $ and $a_j^{\dag}$,  $(j=1,2)$, are the amplitude operators
of the modes. Note that ${\cal V} \in M_4(\mathbb{R})$ is the CM
of a TMGS if and only if the Robertson-Schr\"odinger matrix
inequality holds:
${\cal V}+\frac{i}{2}\Omega\geq 0, \;\;
\Omega:=i(\sigma_2\oplus\sigma_2),$
with $\sigma_2 $ a Pauli matrix. In particular,
${\cal D}:= \det\left({\cal V}
+\frac{i}{2}\Omega \right)\geq 0.$
Gaussian states whose CMs are connected by local symplectic transformations have
 the same {\em amount of entanglement}
and belong to an equivalence class: their CMs are locally
congruent to CMs having a {\em scaled standard form},
\begin{eqnarray}
{\cal V}(u_1,u_2)=\left(\begin{array}{cccc}b_1 u_1 &0&
c\sqrt{u_1u_2} &0\\
0& b_1/u_1&0&d/\sqrt{u_1u_2}\\ c\sqrt{u_1u_2}  &0&b_2 u_2&0\\0
&d/\sqrt{u_1u_2}&0&b_2/u_2 \end{array}\right). \label{tri}
\end{eqnarray}
In Eq.\ (\ref{tri}), $u_1\geq 1, u_2\geq 1$ are one-mode squeezing factors. The unscaled standard form  ${\cal V}(1,1)$ of the CM, introduced in Ref.\cite{Duan}, is expressed in terms
of four parameters $b_1, \; b_2, \;c, \;d$. They are local invariants and determine the entanglement properties of the whole equivalence class.
Recall the locally invariant Peres-Simon separability condition
for a TMGS \cite{Simon},
$ \tilde{\cal V}+\frac{i}{2}\Omega\geq 0,$ with $ \tilde{\cal V}$
denoting the CM of the partially transposed density operator.
This matrix inequality  reduces to the Simon separability test
\cite{Simon}:
\begin{equation}
\tilde{\cal D}:={\rm det}\left(\tilde{\cal V}
+\frac{i}{2}\Omega \right)=\det{\cal V}-\frac{1}{4}(b_1^2+b_2^2
+2 c |d|)+\frac{1}{16}\geq 0.
\label{Sp2}
\end{equation}
The concept of classicality (existence of the Glauber-Sudarshan $P$
representation of the density operator)  is central in our present treatment of the EF. A TMGS with a CM \ (\ref{tri}) is classical
if and only if the matrix ${\cal V}(u_1,u_2)-\frac{1}{2}I_4$
is non-negative, with $I_4$ the $4\times 4$ identity matrix.
This requirement is equivalent to the non-negativity
of all its principal minors. Remark that the classicality conditions are not
locally invariant, depending on the factors $u_1,u_2$.

We start on the programme of Eq.\ (\ref{EF}) for an inseparable mixed TMGS $\rho_G,$ whose CM has a scaled standard form\ (\ref{tri}).
Its four standard-form parameters $b_1, b_2, c\geq |d|=-d>0$
are given, while the scaling factors $u_1,u_2$ are unknown.
Continuous pure-state decompositions of such a state  are convex combinations of the type
\begin{equation}   \rho_G=\int{\rm d}^{2}\beta_1 {\rm d}^{2}\beta_2
P(\beta_1,\beta_2)|\Psi(\beta_1,\beta_2)\rangle
\langle\Psi(\beta_1,\beta_2)|.  \label{2.1}
\end{equation}
$P(\beta_1,\beta_2)$ is a non-negative normalized distribution
function and $|\Psi(\beta_1, \beta_2)\rangle $ is a state vector depending on the complex variables $\beta_1,\beta_2$.
In accordance with the EF definition, Eq.\ (\ref{EF}),
the pure states in the above continuous combination should
achieve an optimal decomposition of the given state $\rho_G$.
To this end, we make use of an important theorem regarding
the ranking of entanglement among pure two-mode states proven
in a recent paper of Giedke {\em et al.} \cite{Giedke}:
For a given EPR uncertainty, the minimal entanglement 
over the whole class of pure states is reached by a Gaussian one,
the two-mode squeezed vacuum state (TMSVS). This important result
leads to the {\em key idea} of our treatment: Owing to the Gaussian
nature of the two-mode state $\rho_G$, as well as to the scaled standard form\ (\ref{tri}) of its CM, we are allowed
from the very beginning to restrict ourselves in Eq.\ (\ref{2.1})
to equally entangled pure states obtained by displacing
a unique TMSVS. Among all ensembles of such pure two-mode states
that realize the given mixed state $\rho_G$ we have to find
the one possessing the minimal entanglement. Let us denote
by $\rho_{0}= |\Psi_0\rangle \langle \Psi_0|$ the TMSVS
entering this optimal convex expansion:
\begin{equation}
{\rho_G}=\int{\rm d}^{2}\beta_1 {\rm d}^{2}\beta_2
P(\beta_1,\beta_2)D_1(\beta_1)D_2(\beta_2)\rho_{0}
D_2^{\dag}(\beta_2)D_1^{\dag}(\beta_1). \label{dec}
\end{equation}
According to Eq.\ (\ref{dec}), the exact EF of the given
mixed two-mode state ${\rho_G}$ reduces to the amount
of entanglement of the TMSVS $\rho_0$:
\begin{equation}
E_F({\rho_G})=E(\rho_{0}). \label{s}
\end{equation}
Recall now that a TMSVS is a Gaussian state whose CM has
precisely the unscaled  standard form  ${\cal V}(1,1)$,
Eq.\ (\ref{tri}): its parameters  $b_1=b_2=:x>1/2, c=-d=:y>0$
are subjected to the purity condition
\begin{equation}
x^2-y^2=\frac{1}{4}. \label{xy}
\end{equation}
The entanglement of a TMSVS is the von Neumann entropy
of its one-mode reductions,
\begin{equation} E (\rho_0)=(x+\frac{1}{2})\ln (x+\frac{1}{2})
-(x-\frac{1}{2})\ln (x-\frac{1}{2}), \label{ent}
\end{equation}
which is an increasing and concave function of the variable
$x>\frac{1}{2}$.
We notice that the optimal convex decomposition\ (\ref{dec})
stands the pertinent test of the pure-state limit case:
$\rho_G=\rho_0$, with $u_1=u_2=1$ and
$P(\beta_1,\beta_2)=\delta_2(\beta_1)\delta_2(\beta_2)$.
To evaluate the EF\ (\ref{s}), one has to be able to determine
the optimal decomposition  \ (\ref{dec}), {\em i.e.}, both the distribution function $P(\beta_1,\beta_2)$ and the TMSVS $\rho_0$. Provided that this can be effectively done for any TMGS,
Eq.\ (\ref{dec}) displays the first result of our work:
{\em the EF for a two-mode Gaussian state coincides with its GEF}.

We now take advantage of a fact well known in quantum optics: decompositions of the type \ (\ref{dec}) do have a clear meaning  starting with Glauber's seminal work on the coherent states
of the electromagnetic field  \cite{G63}. Accordingly,
Eq.\ (\ref{dec}) gives the density operator ${\rho_G}$
of a superposition of two fields: one is in a classical state
$\rho_C$ having the regular Glauber-Sudarshan
$P$ representation $P(\beta_1,\beta_2)$, and the other is
in the pure state  $\rho_{0}$. By employing the corresponding CFs
and writing the $P$ representation as the Fourier transform
of the normally-ordered CF,
$\chi^{(N)}(\lambda_1,\lambda_2):=\exp{\left(\frac{1}{2}
(|\lambda_1|^2+|\lambda_2|^2)\right)}\chi (\lambda_1,\lambda_2)$,
Eq.\ (\ref{dec}) leads to the multiplication law
\begin{equation}
\chi_G^{(N)}(\lambda_1,\lambda_2)=\chi_0^{(N)}(\lambda_1,\lambda_2)
\chi_C^{(N)} (\lambda_1,\lambda_2), \label{3chi}
\end{equation}
with $\chi_C$ denoting the CF of the classical state $\rho_C$ of the superposed field \cite{MM,com}. It follows that the CF
$\chi_C (\lambda_1,\lambda_2)$  is also
Gaussian. Equation \ (\ref{3chi}) results in an addition rule
for the CMs of the Gaussian states involved:
\begin{equation}
{\cal V}(u_1,u_2)={\cal V}_0+{\cal V}_C-\frac{1}{2}I_4. \label{3v}
\end{equation}

Our method towards finding the optimal pure-state decomposition
concentrates on the properties of the classical state
${\rho}_{C}.$  We first show that ${\rho}_{C}$ belongs to the
boundary $\partial {\cal P}$ of the set ${\cal P}$ of all classical TMGSs ({\em Property 1}). Then we prove that ${\rho}_{C}$
is also on the boundary  $\partial {\cal S}$ of the larger set
${\cal S}$ of all separable TMGSs: ${\cal S} \supset {\cal P}$
({\em Property 2}), Refs. \cite{Eng,Oliv}.

{\em Property 1: The superposed classical state $\rho_C$ is at the classicality threshold.}
 Indeed, for the optimal superposition, the CM  ${\cal V}(u_1,u_2)$ should be as close as possible to ${\cal V}_0$. This happens when
the principal minors of rank 3 and 4 of the non-negative matrix
${\cal V}(u_1,u_2)-{\cal V}_0={\cal V}_C-\frac{1}{2}I_4$  are zero.
By the same token, the Gaussian state $\rho_C$ is at the border
of classicality $\partial {\cal P}$. Explicitly, the condition
$\det({\cal V}_{C}-\frac{1}{2}I_4)=0$ holds with the left-hand side
expressed as a product of two vanishing factors:
\begin{eqnarray}(b_1 u_1-x)(b_2 u_2-x)-(c\sqrt{u_1u_2}-y)^2=0,
\label{c1}
\end{eqnarray}
\begin{eqnarray}(b_1/ u_1-x)(b_2/ u_2-x)-
(|d|/\sqrt{u_1u_2}-y)^2=0. \label{c2}
\end{eqnarray}
Equations \ (\ref{c1}) and \ (\ref{c2}) are in agreement with
the Gaussian optimality conditions written in the pioneering work
Ref.\cite{Wolf} on different grounds.
 Making use of Eqs.\ (\ref{xy}), \ (\ref{c1}),
and \ (\ref{c2}), we can impose to the one-variable function
 $x=x(u_1, u_2(u_1))$ the minimization condition
$\frac{{\rm d} x}{{\rm d} u_1}=0.$
We get therefore a fourth independent algebraic equation,
\begin{eqnarray}\frac{b_1 u_1-x}{b_1/ u_1-x}=\frac{b_2 u_2-x}
{b_2/ u_2-x}, \label{c3}
\end{eqnarray}
which implies an additional property of the state
$\rho_C.$

{\em Property 2: The superposed classical state $\rho_C$ is
at the separability limit as well.}
To prove this statement, we use Eqs.\ (\ref{xy}), \ (\ref{c1}),
and \ (\ref{c2}) to evaluate the Simon invariant $\tilde{\cal D}$,
Eq. \ (\ref{Sp2}), of the Gaussian state $\rho_C$.
Taking into account Eq.\ (\ref{c3}), we get $\tilde{\cal D}=0,$
{\em i. e.}, $\rho_C \in \partial {\cal S}.$

The evaluation of the required EF reduces to solving
a system of four non-linear algebraic equations, namely,
Eqs.\ (\ref{xy}), and\ (\ref{c1})--\ (\ref{c3}), with four unknowns:
$u_1,u_2, x, y.$ Let us denote its solution by $w_1,w_2, x_m, y_m.$
The above algebraic system yields a quartic equation, $
\sum_{n=0}^{4} {\cal A}_n p^n=0$,
for the product $p:=u_1 u_2$.
The coefficients  ${\cal A}_n$ are quite simple polynomials in the four
standard-form parameters of the given inseparable TMGS:
\begin{eqnarray*}{\cal A}_0&=&(b_1 b_2-d^2) \left[b_1(b_1 b_2-d^2)-\frac{b_2}{4}\right]\nonumber\\ && \times
\left[b_2(b_1 b_2-d^2)-\frac{b_1}{4}\right]>0,\end{eqnarray*}
\begin{eqnarray*}{\cal A}_1&=&-[c(b_1 b_2-d^2)+\frac{|d|}{4}]
\left\{(b_1-b_2)^2[c(b_1 b_2-d^2)
+\frac{|d|}{4}]\right.\nonumber\\ && \left.
+2b_1 b_2 (c-|d|)\left(b_1 b_2-d^2-\frac{1}{4}\right)\right\}
\leq 0,\end{eqnarray*}
\begin{eqnarray*}
{\cal A}_2&=&
[(b_1 c-b_2|d|)(b_1|d|-b_2 c)+c |d| {\cal Z} ](\det{\cal V}+{1}/{16})\nonumber\\ &&
-2(b_1^2b_2^2-c^2d^2){\cal D}
-c|d|\det{\cal V},
\end{eqnarray*}
\begin{eqnarray}
{\cal A}_3={\cal A}_1(c\leftrightarrow |d|),\;{\cal A}_4
={\cal A}_0(c\leftrightarrow |d|) \geq 0. \label{coef}
\end{eqnarray}

We have introduced the symplectic invariant ${\cal Z}:=b_1^2+b_2^2+2 c d \geq 1/2$.
At $p=1$, the above quartic polynomial has a negative value, except for $c=|d|$, when it vanishes. This implies the existence of a convenient root $p_m=w_1 w_2\geq 1$ for any inseparable mixed TMGS. Had we got $p_m$, it could be used
to obtain the optimal $y_m$ as the smallest root of a quadratic trinomial ${\cal B}_2(p)y^2+{\cal B}_1(p)y+{\cal B}_0(p)$  whose coefficients,

$${\cal B}_0(p)=-\tilde{\cal D}p \geq 0,$$
\begin{eqnarray*}{\cal B}_1(p)&=&-2\sqrt{p}\left(\left[|d|(b_1 b_2-c^2)
+{c}/{4}\right]p\right. \nonumber \\ && \left.+\left[c(b_1 b_2-d^2)
+|d|/{4}\right]\right) <0,\end{eqnarray*}
 \begin{eqnarray}
{\cal B}_2(p)=(b_1 b_2-c^2)p^2+ {\cal Z}p
+(b_1 b_2-d^2) >0,
\label{trin}
\end{eqnarray}
are evaluated at $p=p_m$.
We mention that in four significant particular cases (defined by special relations between standard-form parameters) we have found simple solutions by direct use of Eqs.\ (\ref{c1})--\ (\ref{c3}).
We have then recovered them  by exploiting
Eqs.\ (\ref{coef}) and \ (\ref{trin}).

As a first salient example, we consider an entangled symmetric TMGS, whose standard-form parameters are $b_1=b_2=:b,\;\;c \geq |d|=-d>0$. The smallest symplectic eigenvalue  $\tilde{\kappa}_-$ of the CM for the partially transposed density operator is in this case
$\tilde{\kappa}_-=\sqrt{(b-c)(b-|d|)}.$
In agreement with the results of the remarkable work
Ref.\cite{Giedke},
Eqs.\ (\ref{c1})--\ (\ref{c3}) and \ (\ref{xy}) give:
$$ w_1=w_2=\sqrt{\frac{b-|d|}{b-c}},\;\;\;
x_m=\frac{\tilde{\kappa}_-^2+1/4}{2 \tilde{\kappa}_-}.$$

A second class of notable bipartite states is that
of two-mode squeezed thermal states. The standard-form parameters
of such a state are $b_1 \geq b_2,\;\;c=|d|=-d>0$. This case
was considered previously in Refs. \cite{J,AI}, where
the prescription of Ref.\cite{Wolf} to evaluate the GEF was followed.
From our results, $$w_1=w_2=1,\;\; x_m=\frac{(b_1+b_2)
(b_1b_2-c^2+1/4)-2 c\sqrt{{\cal D}}}{(b_1+b_2)^2-4 c^2},$$
one can see that $x_m$ is not determined only by the eigenvalue
 $\tilde{\kappa}_-=\frac{1}{2}[b_1+b_2
-\sqrt{(b_1-b_2)^2+4c^2}].$

A third example is that of a TMGS at the separability boundary:
$\tilde{\cal D}=0 \Longleftrightarrow \tilde{\kappa}_-=\frac{1}{2}.$
We get
$ x_m=\frac{1}{2}, \;\;\;y_m=0 $ and the optimal squeeze factors
$$w_1=\sqrt{\frac{b_2(b_1b_2-d^2)-\frac{1}{4}b_1}
{b_2(b_1b_2-c^2)-\frac{1}{4}b_1}},
\;\; w_2= w_1(b_1\leftrightarrow b_2).$$
Equation \ (\ref{dec}) becomes now the $P$ representation
of a state $\rho_G$ at the border of classicality
$\partial {\cal P}$ and that of separability $\partial {\cal S}$
as well.

A fourth class of entangled states consists of those TMGSs
whose CMs have the smallest symplectic eigenvalue $\kappa_-$:
${\cal D}=0 \Longleftrightarrow {\kappa}_-=\frac{1}{2}.$
These states were studied as having minimal negativity
at fixed local and global purities \cite{Aa}. Assuming that
$b_1\geq b_2,\;\;c\geq |d|=-d>0$, we found two distinct solutions  required by the sign of the difference $b_2 c-b_1|d|$:
$$b_2 c-b_1|d| < 0: \;\;\;x_m=\frac{b_1^2-b_2^2}{8(\det {\cal V}-\frac{1}{16})},$$
$$w_1=\sqrt{\frac{b_2(b_1b_2-d^2)-\frac{1}{4}b_1}
{b_2(b_1b_2-c^2)-\frac{1}{4}b_1}},\;\;
w_2=w_1(b_1\leftrightarrow b_2).$$
$$b_2 c-b_1|d|\geq 0:\;\;
x_m=\frac{1}{2}\sqrt{\frac{b_1b_2}{b_1b_2-d^2}},$$
$$w_1=2\sqrt{\frac{b_1}{b_2}(b_1b_2-d^2)},\;\;
w_2=w_1(b_1\leftrightarrow b_2).$$

The above formulae for $x_m$ are
in agreement with those derived in other parametrization
in Ref.\cite{AI}, which follows the methods of Ref.\cite{Wolf}.

The last issue we are here interested in is the additivity
of the EF for TMGSs. Our present approach gives a straightforward answer to this open question \cite{Plenio}. We consider
a four-mode product state $\rho_G \otimes\sigma_G$,
where $\rho_G$ and $\sigma_G$ are entangled TMGSs. We denote the minimally entangled TMSVSs entering the optimal decompositions of
the type \ (\ref{dec}) for both factors  by $\rho_0$ and $\sigma_0$, respectively.
Therefore,
their tensor product $\rho_0 \otimes \sigma_0$  enters the optimal convex decomposition of the
four-mode state $\rho_G \otimes \sigma_G$.  It follows the identity
$E_F(\rho_G \otimes\sigma_G)=E(\rho_0 \otimes \sigma_0).$
The well-known  additivity property of the von Neumann entropy,
$E(\rho_0\otimes\sigma_0)=E(\rho_0)+E(\sigma_0)$, yields
the additivity of the EF for TMGSs:
$$E_F(\rho_G \otimes\sigma_G)=E_F(\rho_G)+E_F(\sigma_G).$$
Consequences of this property on evaluating other measures of entanglement are largely discussed in Ref.\cite{Plenio}.

To sum up, we have reformulated the problem of evaluating the EF
for TMGSs in terms of CFs and CMs. We have shown that the exact EF
of such a state coincides with its Gaussian one. Although
an analytic solution in the general case seems to be complicated,
it can be found, nevertheless, by solving a quartic equation.
Our general treatment allowed us to retrieve readily
previous explicit results in some relevant particular cases.
Based on our approach, we have finally proven the additivity
of the EF for two-mode Gaussian states.

This work was supported  by the Romanian Ministry of Education 
and Research through Grant No. IDEI-995/2007 
for the University of Bucharest.

{\em Note added}.-During the completion of this Letter, an interesting
treatment of the EF for a TMGS was given in Ref.\cite{Simon1}.
Its relation to our present work will be discussed elsewhere.


\begin{thebibliography}{99}
\bibitem{com} Equation\  (\ref{3chi}) can be obviously generalized
to arbitrary multimode states. The only validity condition
is the classicality of the state $\rho_{C}$. In Ref.\cite{G63},
only superpositions of classical fields were considered
and the convolution law of the corresponding well-behaved
$P$ representations was derived. In our paper \cite{MM},
we pointed out the multiplication law of the CFs and employed it
to analyze the influence of thermal noise on nonclassical
properties of one-mode states.
\end{thebibliography}
\end{document}